\documentstyle[prl,preprint,aps]{revtex}
\narrowtext
\begin{document}

\title{Cluster Monte Carlo Simulations of the Nematic--Isotropic Transition}
\author{N. Priezjev and Robert A. Pelcovits}
\address{Department of Physics, Brown University, Providence, RI 02912}
\date{\today}
\maketitle

\begin{abstract}

We report the results of simulations of the Lebwohl-Lasher model of the nematic--isotropic transition using a new cluster Monte Carlo algorithm.  The algorithm is a modification of the Wolff algorithm for spin systems, and greatly reduces critical slowing down. We calculate the free energy in the neighborhood of the transition for systems up to linear size 70.  We find a double well structure with a barrier that grows with increasing system size, obeying finite size scaling for systems of size greater than 35. We thus obtain an estimate of the value of the transition temperature in the thermodynamic limit.
\end{abstract}

\pacs{64.70.Md, 61.30.Cz, 87.53.Wz}
\pagestyle{headings}

The Lebwohl--Lasher (LL) model \cite{Lasher:72} is a lattice model of rotors with an orientational order--disorder transition. The long axes of the rotors are specified by unit vectors 
${\mathbf \sigma}_i$. While it neglects 
the coupling between the orientational and translational degrees of freedom present in 
a real nematic liquid crystal, it is generally believed that this coupling does not play 
a significant role at the nematic--isotropic (NI) transition. With the absence of translational 
degrees of freedom the LL model is particularly well--suited for large--scale simulations 
of the transition. The model is defined by the Hamiltonian: 

\begin{equation}
{\cal {H}}= -\epsilon \sum _{<ij>}\biggl\lbrace {3 \over 2}
({\mathbf \sigma}_{i} \cdot  {\mathbf \sigma}_{j})^2 -{1\over 2}\biggr\rbrace
\end{equation}
where the sum is over all nearest-neighbors and $\epsilon$ is a coupling parameter. The LL model has been intensively investigated using Monte Carlo techniques since its introduction \cite{Luckhurst,Fabbri,Biscarini,Zannoni,Zhang:93,Boschi}. 

As in real experimental systems 
the NI transition in the LL model is weakly first--order; thus, there is significant critical slowing down in the neighborhood of the transition. In a Monte Carlo simulation the system gets trapped in one of the free energy wells corresponding to 
the nematic or isotropic phase, and the conventional single flip 
Metropolis algorithm becomes inefficient especially as the system size is increased.  While Boschi {\it et al.}~\cite{Boschi} carried out simulations on systems as large as $120 \times 120 \times 120$, the most detailed study of the NI transition was carried out by Zhang {\it et al.}~ \cite{Zhang:93} on systems up to $28 \times 28 \times 28$. These authors used the Lee--Kosterlitz finite size scaling method \cite{Kosterlitz:90,Kosterlitz:91}, supplemented by the Ferrenberg--Swendsen reweighting technique \cite{Ferrenberg:89} to determine the order of the NI transitions and estimate the value of the transition temperature $T_c$ in the thermodynamic limit. In the Lee--Kosterlitz method one examines the finite size scaling of the free energy barrier $\Delta F$ between the nematic and isotropic phases; at a first--order transition this should be an increasing function the linear system size $L$, while it should approach a constant for systems with continuous phase transitions. For a large enough system, specifically $L \gg \xi$, where $\xi$ is the correlation length, a finite size scaling analysis predicts that $\Delta F \sim L^2$ for three--dimensional systems. In the LL model Zhang {\it et al.}~found a small free energy barrier appearing at the two largest system sizes they studied, $L=24$ and 28, and thus did not have enough data to carry out a finite size scaling analysis of $\Delta F$. Instead they estimated the value of $T_c$ in the thermodynamic limit by extrapolating three different measures of $T_c$: the positions of the maxima in the specific heat and susceptibility and the the temperature where the two free energy wells are of equal depth. However, as we demonstrate below, the system sizes considered by Zhang {\it et al.}~are not in the finite--size scaling regime, and thus their estimate of $T_c$ in the thermodynamic limit is not accurate. 

Over the past decade significant advances have been made in algorithm development which 
overcome critical slowing down in magnetic spin systems \cite{Binder:95}. In particular single cluster algorithms 
have proven to be very efficient in simulating the three--dimensional Ising, XY and Heisenberg 
models. These algorithms are nonlocal updating methods where a single cluster of spins is constructed and the spins within the cluster are updated simultaneously. In the Ising case \cite{SW} clusters of spins are 
formed by creating bonds between parallel spins with a probability that guarantees detailed 
balance. For models with continuous symmetry Wolff \cite{Wolff:89} introduced a cluster algorithm where ``parallel spins''  refer to spins which point in the same hemisphere.  A hemisphere is defined by an equatorial plane perpendicular to a randomly chosen direction $\mathbf{\hat{r}}$. Nematic liquid crystals differ in an important symmetry aspect from magnetic systems, namely, ``up'' and ``down'' spins are 
equivalent.  
To construct a cluster algorithm suitable for simulating the LL model we have modified the Wolff 
algorithm to account for this symmetry difference.
As in the original Wolff algorithm we randomly choose a direction ${\mathbf{\hat{r}}}$. Then we reflect any molecular long axes 
for which ${\mathbf \sigma} _{i} \cdot {\mathbf{\hat{r}} }< 0$, by the transformation 
$\sigma \rightarrow -\sigma$; note that the Hamiltonian ${\cal {H}}$ is invariant 
under this operation. Next we choose a site $i$ at random and reflect it, 
${\mathbf\sigma} _{i}^\prime = R({\mathbf{\hat{r}}}){\mathbf \sigma} _{i}$ using the reflection 
operator $R({\mathbf{\hat{r}}})$ defined by:

\begin{equation}
\label{Rn}
R({\mathbf{\hat{r}}}){\mathbf {\sigma} }_{i} = -{\mathbf {\sigma}} _{i} + 
2 ({\mathbf {\sigma} }_{i} \cdot {\mathbf{\hat{r}}}){\mathbf{\hat{r}}}.
\end{equation}
This operation is illustrated in Fig. \ref{flip}a. Unlike the original Wolff reflection 
operation which reflects spins from the original hemisphere to the opposite one, the present reflection operator keeps the molecular orientation vectors in the same hemisphere defined by 
${\mathbf{\hat{r}}}$. Next we form bonds with the nearest--neighbors of 
${\mathbf {\sigma}} _{i}^\prime$ with probability:

\begin{eqnarray}
\label{Pn}
\ P_{ij} &= &1 - \exp\Biggl\{ {\rm min} \Bigl[ 0, \beta ({\mathbf \sigma}_{i}^\prime \cdot  {\mathbf \sigma}_{j} )^2-\beta ({\mathbf \sigma}_{i}^\prime \cdot (R({\mathbf{\hat{r}}}){\mathbf \sigma}_{j}) )^2  \Bigr] \Biggr\}\nonumber\\ &= &1 - \exp\Biggl\{ {\rm min} \biggl[ 0 , 4 \beta ({\mathbf \sigma}_{i}^\prime \cdot {\mathbf{\hat{r}}})( {\mathbf \sigma}_{j} \cdot {\mathbf{\hat{r}}}) \Bigl(({\mathbf \sigma}_{i}^\prime\cdot {\mathbf \sigma}_{j}) - ({\mathbf \sigma}_{i}^\prime \cdot {\mathbf{\hat{r}}})({\mathbf \sigma}_{j} \cdot {\mathbf{\hat{r}}}) \Bigr)\biggr]\Biggr\},
\end{eqnarray}
where $\beta = 3\epsilon / 2k_BT$. 
This probability is a modification of the one introduced by Wolff, replacing the Heisenberg 
interaction $-J {\mathbf \sigma}_{i} \cdot  {\mathbf \sigma}_{j}$ by the Lebwohl--Lasher interaction 
$-\frac{3}{2}\epsilon({\mathbf \sigma}_{i} \cdot  {\mathbf \sigma}_{j})^2$. As in the original Wolff algorithm we continue this process, forming bonds with the nearest--neighbors of all reflected molecular orientation vectors until the cluster cannot grow any further.

To understand the formation of clusters, consider the projection of two molecular orientation vectors, 
${\mathbf \sigma} _{i}$ and one of its nearest--neighbors ${\mathbf \sigma} _{j}$, on the 
plane perpendicular to ${\mathbf{\hat{r}}}$ before the reflection operation is performed 
(see Fig. \ref{flip}b). 
A bond between these two molecules will likely form if the angle $\phi$ between their projections 
is less than $90^\circ$. 
Note that the probability Eq.(\ref{Pn}) for the bond formation is maximized when the angle between 
${\mathbf \sigma}_{i}^\prime$ and ${\mathbf \sigma}_{j}$ is $90^\circ$ and each of these molecules 
makes an angle of $45 ^{\circ}$ with ${\mathbf{\hat{r}}}$. 
Thus, as in the original Wolff algorithm at low temperatures the molecules are nearly all aligned 
and it is highly probable that large fraction of all molecules will be flipped at once. 
On the other hand at high temperatures the distribution of molecules will be isotropic, 
resulting in flipping small clusters in random directions. In the intermediate region close to the NI
transition temperature the system flips between isotropic and nematic states.

We have used our cluster algorithm to simulate the LL model on a simple cubic lattice of linear dimension $L$, $30\leq L\leq 70$, 
with periodic boundary conditions, in order to study the properties of the NI transition. The temperature $T$ was measured in dimensionless units of 
$ \epsilon/k_B $, in agreement with the units used in previous studies of this model. Our initial random 
configurations were equilibrated at least 200000 Monte Carlo steps (MCS) before starting 
production runs. We found that the average cluster size at 
temperatures close to the NI transition is approximately $0.17 N$ sites per cluster (where $N$ is the total number of lattice sites), essentially independent of system size. Approximately half of the MCS resulted in clusters with fewer 
than ten sites and were efficiently simulated with scalar code. However, a significant fraction 
of clusters had $N/2$ sites or greater, and employing a 
vectorizable cluster construction method \cite{Evertz:93} yielded a sixfold speedup.
\par
For each configuration generated, we calculated the energy per site, $E={\cal {H}}/N$. 
To ascertain the nature of the phase transition we proceeded as in Ref. \cite{Zhang:93} and used the method of Lee and Kosterlitz 
\cite{Kosterlitz:90,Kosterlitz:91}, which relies on the single histogram reweighting technique 
of Ferrenberg and Swendsen \cite{Ferrenberg:89}. Following the approach of the latter authors we  stored 
the configuration data in a histogram $H(E,T,L)$.  The normalized probability distribution 
function $P(E,T,L)$ of the energy is then given by:

\begin{equation}
P(E,T,L)= {H(E,T,L)\over \sum_{E} H(E,T,L)}
\end{equation}
Given this distribution function at temperature $T$, the Ferrenberg--Swendsen method allows the calculation 
of thermodynamic quantities at a different temperature $T^\prime$ in the neighborhood of $T$.  Specifically, thermodynamic quantities at $T^\prime$ can be calculated using the distribution function $P(E,T^\prime,L)$ where: 

\begin{equation}
\label{reweight}
\ P(E,T^\prime,L) = {H(E,T,L)  \exp(-\Delta \beta  E) \over \sum_{E} 
H(E,T,L)\exp(-\Delta \beta  E)} 
\end{equation}
and
\begin{equation}
 \Delta \beta =  (1/T^\prime -1/T). 
\end{equation}
Thus, accurate information over the entire critical region can be extracted from a small number of simulations.

The Lee--Kosterlitz method utilizes the system size dependence of the  barrier $\Delta F$ separating the isotropic and nematic free energy minima at the transition 
temperature to determine the order of the transition. 
If the barrier grows with increasing $L$ then the transition is first order; furthermore, if finite size scaling holds, then $\Delta F \sim L^2$ in a three--dimensional system.
To determine the barrier height we use the free--energy--like quantity :

\begin{equation}
\label{Feqn}
\ F(E,T,L) \sim - \ln P(E,T,L)
\end{equation}
which differs from the true free energy by additive quantities dependent only on $T$ and $L$ 
which are irrelevant to computations of free energy differences. This free--energy--like 
quantity is shown in Fig.\ref{F} for different system sizes, and we note the appearance 
of a pronounced double--well structure for sufficiently large system sizes. 
The right and left hand wells correspond to the nematic and isotropic phases respectively.  In collecting our data we made sure that system made at least 100 hops between the two wells for the largest system size of 70, for a run of 
$6 \times 10^6$  MCS. For each system size, we performed sufficient MCS such that the typical number of points in one 
bin of the histogram $H$ is much larger than the variation in the exponential factor 
$\exp({\Delta T L^3} )$, where $\Delta T = T^\prime - T$. This criterion arises from 
the requirement that the peak in the reweighted distribution Eq. (\ref{reweight}) avoid the 
``wings'' of the measured histogram. Typically we found approximately $10^4$ points in each 
bin and the exponential factor varied by about $10$.

\par
Zhang {\it et al.}~\cite{Zhang:93} made similar plots of the free energy $F(S,T,L)$ as a function of the nematic order parameter $S$ rather than the energy $E$. For the system sizes studied by these authors, with $L \le 28$, the free energy function $F(E,T,L)$ is a much weaker indicator of the nature of the NI transition. However, for the system sizes we have studied, with $30 \le L \le 70$, the free energy as a function of $E$ is a very good indicator as illustrated in Fig.\ref{F}. We calculated $F(S,T,L)$ for $L=28$, the largest system size studied by Zhang {\it et al.}~to check that our cluster algorithm yields the same transition temperature they determined using the conventional single spin flip MC algorithm. In general we did not calculate $F(S,T,L)$ because this requires calculation of a histogram $H(E,S,T,L)$, dependent on $S$ as well as $E$ in order to carry out the reweighting. Calculation of this multiple histogram with sufficiently good statistics is prohibitively time consuming for large systems.
\par
The barrier height $\Delta F(L)$ can be computed as follows :

\begin{equation}
 \Delta F(L) = F(E _{m},T,L) -  F(E _{1},T,L)
\end{equation}
where $E _{m}$ is the energy corresponding to the top of the free energy barrier and
$E _{1}$ is either one of the degenerate local minima. Our results for the barrier height as a function of system size are shown in Fig.\ref{DeltaF}, where we see that finite size scaling holds for systems of size $L \gtrsim 35$, i.e. beyond the largest systems considered in ref. \cite{Zhang:93}. 
The transition temperature $T_c(L)$ for a particular system size $L$ is given by the value of the temperature where 
the two free energy wells have equal depths. Our results for $T_c(L)$ are shown in Fig.\ref{Tc}.
 From the straight line plotted in the figure we see that the finite size scaling relation,
\begin{equation}
\label{Tceqn}
\Delta T = T_c -T_c(L) \simeq L^{-3}.
\end{equation}
works well for systems of size $L \gtrsim 35$, as we would expect, given the scaling of the free energy barrier, $\Delta F(L)$. The intersection of the straight line with the $T_c(L)$ axis yields our estimate of the transition temperature in the thermodynamic limit: $T_c =1.1225 \pm 0.0001$. This value is lower than that obtained in Ref. \cite{Zhang:93}, ($T_{c}=1.1232 \pm 0.0001$).
However, our data indicates that the system sizes studied in Ref. \cite{Zhang:93} were not large enough to carry out a finite size scaling analysis. Even though the three measures of $T_c(L)$ used in this latter reference apparently extrapolate to a single number, there is no justification for using a straight line extrapolation for the transition temperature in the absence of finite size scaling of the free energy barrier.

In conclusion, we have developed a modification of the single cluster Wolff algorithm for nematic systems which has enabled us to efficiently study the NI transition in Lebwohl--Lasher systems of sufficiently large size that finite size scaling is obeyed. As in the case of the cluster algorithms developed originally for ferromagnetic models, our algorithm allows us to overcome the critical slowing down associated with conventional single--flip Monte Carlo. The phase space can then be sampled efficiently near the transition as the system will flip readily between the ordered and disordered phases. We have also applied our algorithm to study the behavior of disclination loops in the transition region \cite{Priezjev} and the efficiency of our algorithm should allow the study of many other interesting properties of the transition.

We thank Prof. J.~M. Kosterlitz for many helpful discussions and suggestions. This work was 
supported by the National Science Foundation under grant DMR--9873849.
Computational work in support of this research was performed at Brown University's 
Theoretical Physics Computing Facility. 

\bibliographystyle{prsty}

\begin{figure} 

\caption{(a) Illustration of the reflection operation $R({\mathbf{\hat{r}}})$, Eq.~(\ref{Rn}). The unit vector ${\mathbf{\hat{r}}}$ is chosen randomly at the start of the algorithm. The reflection operation yields the new molecular orientation ${\mathbf\sigma} _{i}^\prime = R({\mathbf{\hat{r}}}){\mathbf \sigma} _{i}$ as shown. (b) Illustration of the formation of a cluster of two molecules. Here we show the projections of the two molecular long axes ${\mathbf\sigma} _{i}$ and ${\mathbf\sigma} _{j}$ on the plane perpendicular to ${\mathbf{\hat{r}}}$, as well as the projections of the molecular long axes produced by the reflection operator $R({\mathbf{\hat{r}}})$ acting on these two molecules. Two molecules are likely to form a cluster if they each make an angle of approximately $45 ^{\circ}$ with ${\mathbf{\hat{r}}}$ and if the angle $\phi$ between 
their projections is less than $90^\circ$.}

\label{flip}

\end{figure}
\begin{figure}
\caption{Free energy, Eq.~(\ref{Feqn}), in units of $\epsilon$ as a function of the energy per unit site $E$ (also measured in units of $\epsilon$), for four different lattice sizes, $L=30 \ (\bullet), 50 \  (\triangle), 60 \  (\circ), 70 \ (\ast)$. The data for the three largest system sizes have been displaced vertically for the sake of clarity.} 
\label{F}

\end{figure}
\begin{figure}
\caption{The free energy barrier height $\Delta F$ divided by $L$ (measured in units of $\epsilon$ over the lattice spacing) as a function of $L$ (measured in units of the lattice spacing). The straight line fit for system sizes $L \geq 35$ indicates that $\Delta F$ obeys the finite size scaling relation, $\Delta F \sim L^2$, as expected for a first--order transition.} 
\label{DeltaF}

\end{figure}
\begin{figure}
\caption{The transition temperature $T_c(L)$, (measured in units of $\epsilon / k_B$) as a function of $L^{-3}$ (in units of $5\times 10^{-5}$ cubic lattice spacings), for the eight system sizes shown in Fig. \protect\ref{DeltaF}, showing the expected finite size scaling behavior (the straight line fit) given by Eq.~(\ref{Tceqn}) for system sizes $L \geq 35$. The extrapolation of this line to infinite system size yields an estimate of the transition temperature (indicated by the arrow) in the thermodynamic limit.} 
\label{Tc}

\end{figure}

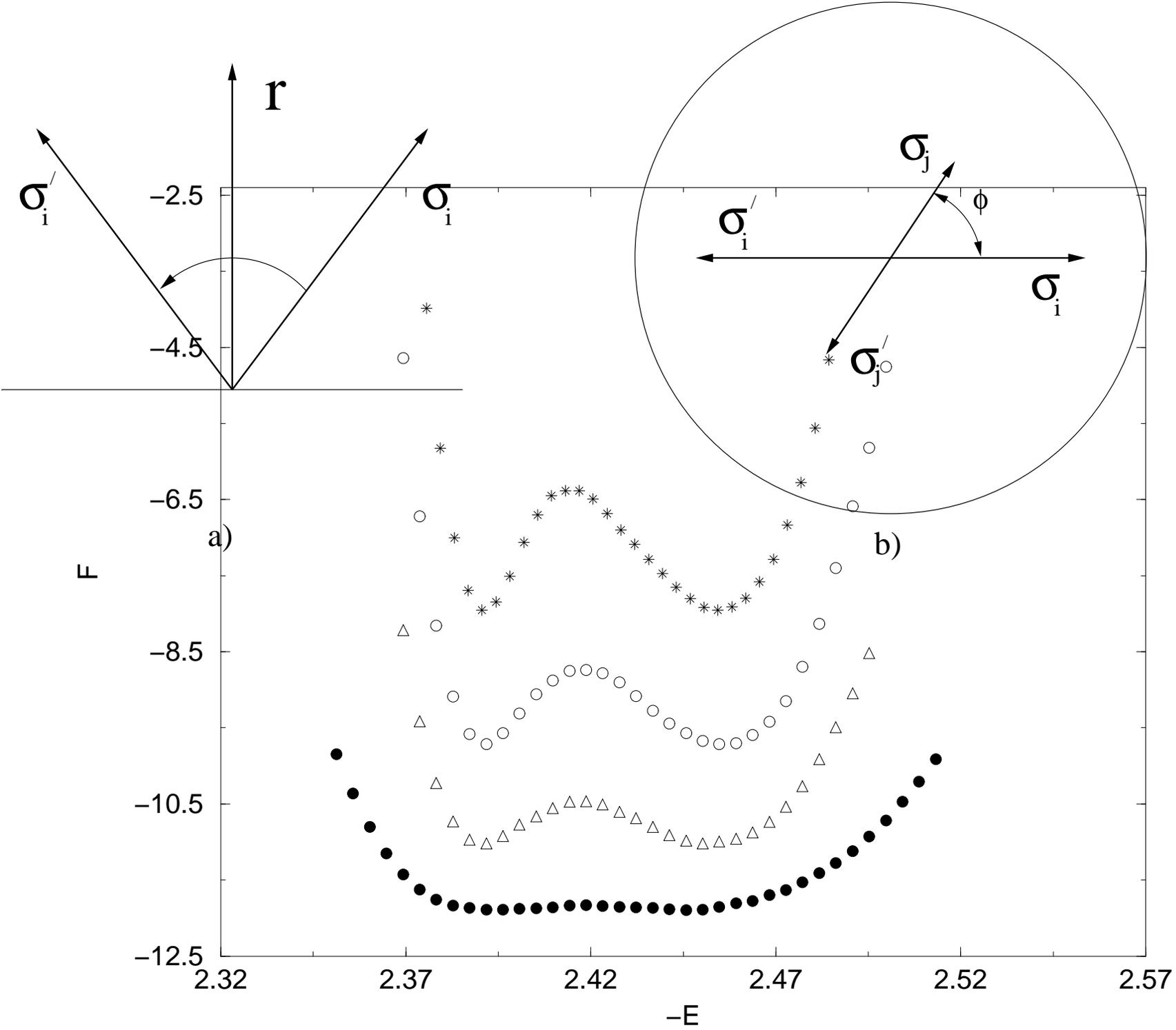
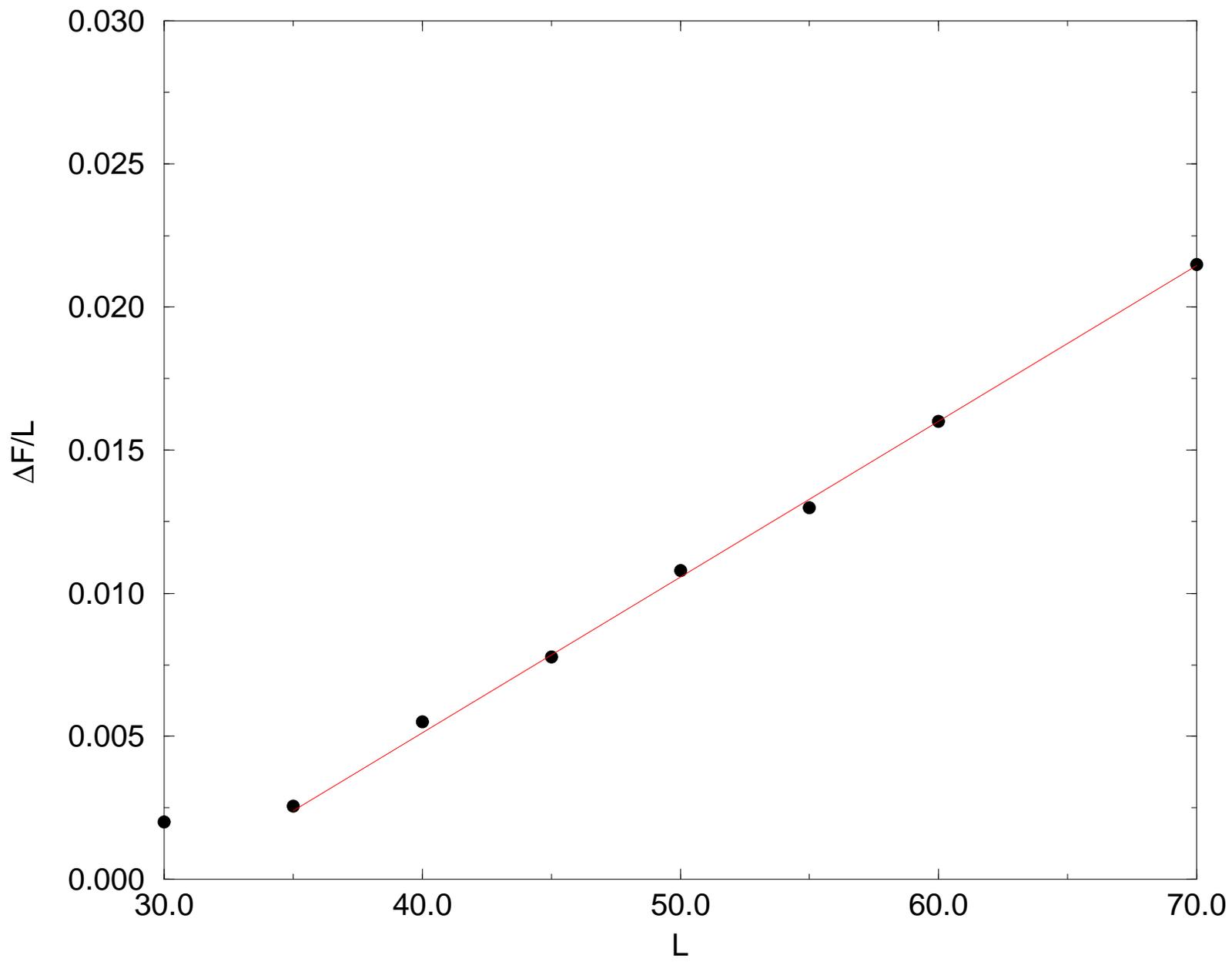
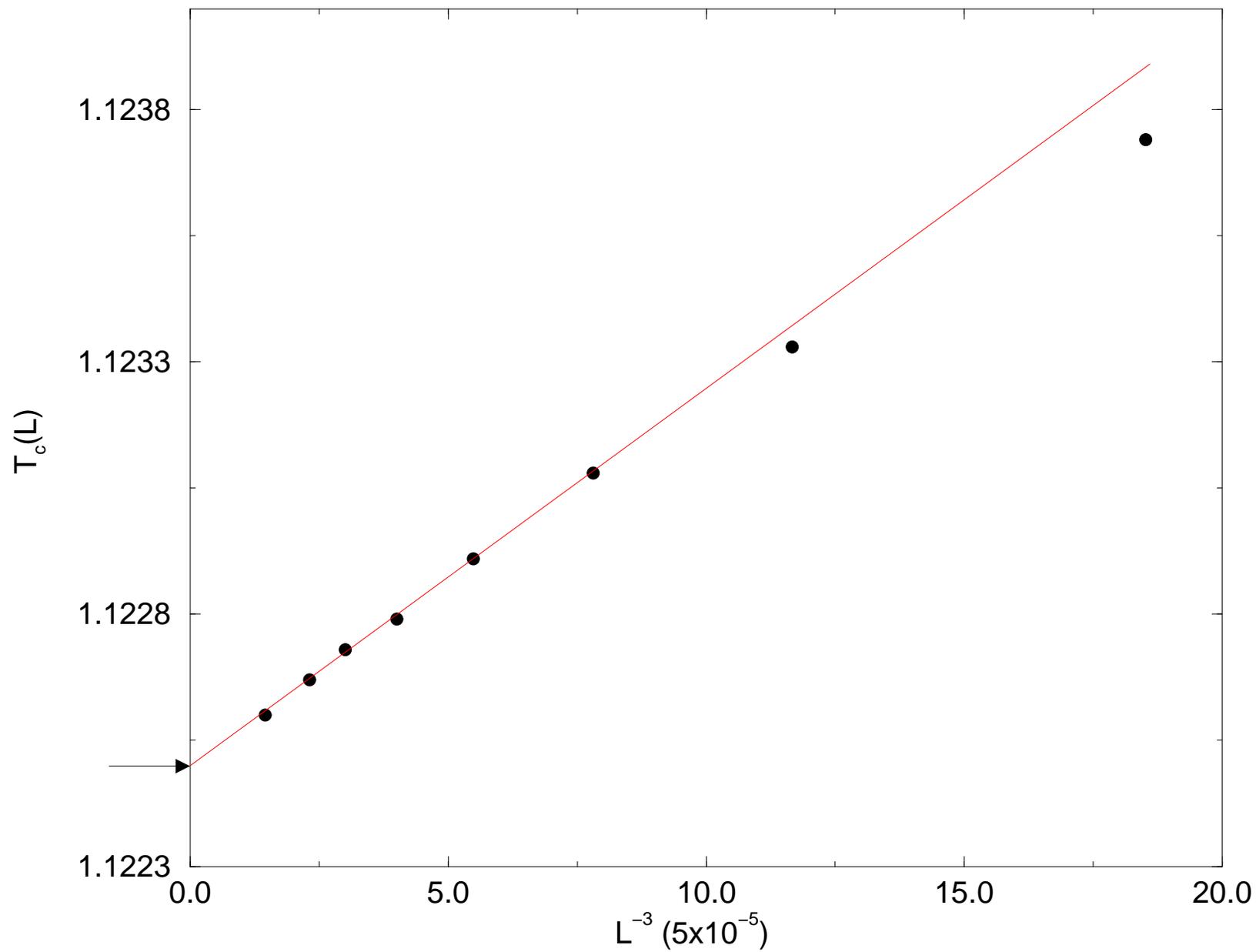
\end{document}